\documentclass[trackchanges, twocolumn]{aastex701}

\usepackage{graphicx}

\usepackage{hyperref} 

\usepackage{amstext}
\usepackage{tikz}
\usetikzlibrary{graphs, positioning, quotes, shapes.geometric}
\usepackage{mathrsfs}
\usepackage{mhchem}
\usepackage{orcidlink}
\usepackage{threeparttable}

\usepackage{txfonts}

\begin{document}

\title{First mapping of prebiotic molecule \ce{CH2NH} in a pre-stellar core}

\author[orcid=0000-0001-9299-5479]{Yuxin Lin}
\affiliation{Max-Planck-Institut f{\"u}r Extraterrestrische Physik, Giessenbachstr. 1, D-85748 Garching bei M{\"u}nchen}
\email[show]{ylin@mpe.mpg.de}  

\author{Silvia Spezzano} 
\affiliation{Max-Planck-Institut f{\"u}r Extraterrestrische Physik, Giessenbachstr. 1, D-85748 Garching bei M{\"u}nchen}
\email{spezzano@mpe.mpg.de}

\author{Olli Sipil\"a}
\affiliation{Max-Planck-Institut f{\"u}r Extraterrestrische Physik, Giessenbachstr. 1, D-85748 Garching bei M{\"u}nchen}
\email{spezzano@mpe.mpg.de}

\author{Jaime E. Pineda}
\affiliation{Max-Planck-Institut f{\"u}r Extraterrestrische Physik, Giessenbachstr. 1, D-85748 Garching bei M{\"u}nchen}
\email{jpineda@mpe.mpg.de}

\author{Paola Caselli}
\affiliation{Max-Planck-Institut f{\"u}r Extraterrestrische Physik, Giessenbachstr. 1, D-85748 Garching bei M{\"u}nchen}
\email{caselli@mpe.mpg.de}
\begin{abstract}

We present the first spatially resolved map of methanimine (\ce{CH2NH}) in the prestellar core L1544 using the IRAM 30m telescope. The 2$_{0,2}$--1$_{0,1}$ line at 127 GHz was mapped with 20\arcsec~resolution ($\sim$2800 au), revealing extended \ce{CH2NH} emission across the core. The peak line intensity coincides with the well-known \ce{c-C3H2} peak, while the integrated intensity peaks between the HNCO and dust continuum peaks due to broader linewidths in the latter region. Column densities of \ce{CH2NH} are $\sim$(0.5--1.4$\times$)10$^{12}$ cm$^{-2}$, corresponding to fractional abundances of $5\times10^{-11}$--$1\times10^{-10}$, with a trend decreasing from the southern, carbon-chain rich region to the dust and HNCO peak in the north. Comparison with complementary molecular maps and the gas-grain chemical model of Sipil\"a et al. suggests that neutral--neutral gas-phase reactions and dissociative recombination dominate in the outer carbon-chain shell. This study demonstrates that \ce{CH2NH}, a simple nitrogen- and carbon-bearing molecule previously detected with pointed observations in other cold cores, is present and spatially extended in the evolved pre-stellar core L1544. This indicates that prebiotic nitrogen–carbon chemistry continues efficiently up to the onset of gravitational collapse, providing key constraints for astrochemical models and the early stages of chemical complexity leading to amino acids.

\end{abstract}


\section{Introduction}

\ce{CH2NH} is the simplest interstellar imine and a key nitrogen-bearing organic molecule that serves as a building block for more complex prebiotic species. Laboratory experiments and astrochemical models identify \ce{CH2NH} as a critical intermediate in the formation of amino acids \citep{Danger11, Joshi2022}, particularly glycine (\ce{NH2CH2COOH}), the simplest amino acid, consisting of an amino group (-\ce{NH2}), a carboxyl group (-\ce{COOH}), and a single hydrogen atom as its side chain. Owing to this minimal structure, glycine is the only achiral amino acid and represents the basic framework of all $\alpha$-amino acids. Imine species play a central role in nitrogen chemistry as intermediates linking nitriles and amines, thereby bridging simple N-bearing molecules to biologically relevant compounds. In interstellar ices, \ce{CH2NH} may form via partial hydrogenation of \ce{HCN} \citep{Theule11} or through UV- and cosmic-ray–induced processing of methanimine-bearing ices \citep{Zhu19, Keresztes24}, though the relative efficiency of these routes remains uncertain.

The detection of \ce{CH2NH} in a variety of astrophysical environments highlights its relevance to prebiotic chemistry: \ce{CH2NH} has been observed in hot massive cores or hot corinos \citep[e.g., Orion-KL, Sgr B2, IRAS16293-2422B;][]{Godfrey73, Dickens97, Suzuki16, Ligerink18}, cold dark clouds \citep[e.g., L183, Ori-3N;][]{Turner99, Dickens97}, circumstellar envelopes \citep{Tenenbaum10}, Titan \citep{Vuitton06}, and even external galaxies (\citealt{Salter08}; \citealt{Gorski21}). These detections show that N--C-bearing prebiotic species can form and survive across a wide range of physical conditions, from cold, quiescent clouds to regions with intense star formation.

Compared to the frequent detection of \ce{CH2NH} in massive star-forming regions and more extreme environments, detections of \ce{CH2NH} in low-mass objects remain few. A sensitive Atacama Large Millimeter Array (ALMA) broad line survey toward IRAS16293-2422B, a reknown hot corino, confirmed the first detection of \ce{CH2NH} in such objects \citep{Ligerink18}. In even earlier evolutionary stages, before protostar formation, detections are scarce \citep{Margules22}: only L183 and Ori-3N show confirmed emission (pointed observations), while in the carbon-chain–rich core TMC-1, only upper limits have been reported \citep{Turner99, Luthra23}. 

In warm regions, models suggest that \ce{CH2NH} forms efficiently through gas-phase neutral–neutral reactions (e.g., \ce{CH3 + NH}) following ice sublimation during the warm-up phase \citep{Suzuki16}. In contrast, its origin in cold, dense environments remains poorly understood. The observed abundance in L183 ($\sim8\times10^{-10}$) exceeds gas-phase model predictions by one to two orders of magnitude, implying additional formation pathways such as grain-surface chemistry, cosmic-ray–induced photochemistry, or ion–molecule reactions \citep{Luthra23}. Understanding the balance between these mechanisms requires spatially resolved observations across well-characterized environments that bridge gas-phase and grain-surface chemical regimes.

Here we present the first resolved mapping of \ce{CH2NH} emission in L1544, a well-studied pre-stellar core characterized by a dense, heavily depleted center ($n\sim10^6$~cm$^{-3}$; \citealt{KC10, Caselli22}) and a chemically rich and differentiated envelope \citep{Spezzano16, Spezzano17}. We compare the observed abundance and spatial distribution of \ce{CH2NH} with one-dimensional chemical models to investigate its dominant formation routes. This study provides new insights into the emergence and spatial organization of prebiotic molecules in cold, dense gas.

\begin{figure}
    \includegraphics[width=0.53\textwidth]{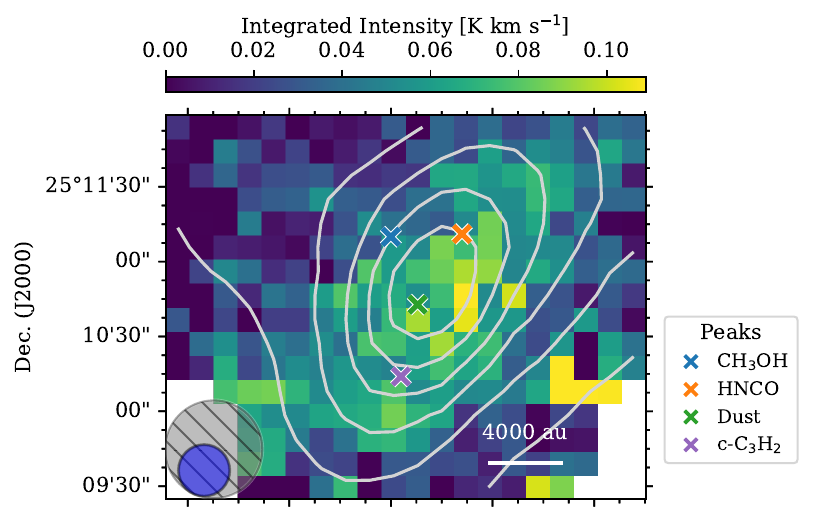}\\
    \hspace{2cm}\includegraphics[width=0.5\textwidth]{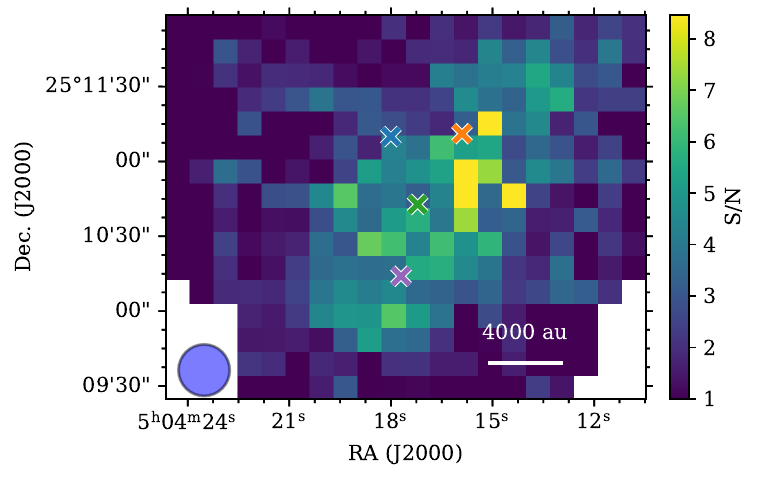}\\
   \hspace{3cm}\includegraphics[width=0.42\textwidth]{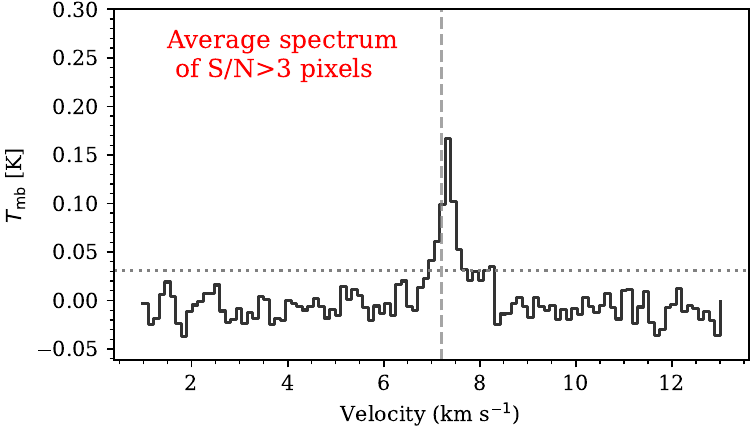}
    \caption{Top: Integrated intensity map of \ce{CH2NH} 2$_{\rm{0,2}}$-1$_{\rm{0,1}}$ line toward L1544. The velocity range for integration is 6.8-7.6~km~s$^{-1}$ (system velocity, $V_{\mathrm{LSR}}$ = 7.2~km~s$^{-1}$). Gray contours indicate molecular hydrogen column density ($N_{\mathrm{H2}}$) derived from {\it{Herschel}} (\citealt{Spezzano16}) from 3.0$\times$10$^{20}$-2.7$\times$10$^{22}$~cm$^{-2}$ with six uniform intervals. The three molecular peaks and the dust peak are indicated with cross markers (\citealt{Spezzano17}). Beam sizes of the line (20$''$, in blue) and of $N_{\mathrm{H2}}$ map (38$''$, in gray) are indicated in the bottom left corner. Middle: Signal-to-noise (S/N) ratio map based on integrated intensity maps divided by $\sigma_{\rm rms} \times \sqrt{N_{\rm chan}} \times \Delta V$; markers follow same definition as in the top panel. Bottom: Average spectrum from S/N$>$3 pixels; $V_{\mathrm{LSR}}$ is shown as a vertical dashed line and 2$\sigma$ noise level as a horizontal dotted line.}
    \label{fig:mom0}
\end{figure}

\section{Observation and data reduction}

The \ce{CH2NH} 2$_{0,2}$--1$_{0,1}$ line at 127.857~GHz (Table \ref{tab:line_info}) was mapped toward the pre-stellar core L1544 with the IRAM 30m telescope (Pico Veleta, Spain) during winter semester of 2024 (project ID: 093-24, PI: Yuxin Lin). Observations were performed in on-the-fly (OTF) mapping mode under good weather conditions ($\tau$$\sim$0.2) using the EMIR E150 receiver in combination with the Fourier Transform Spectrometer (FTS50) backend, providing a spectral resolution of 50~kHz ($\sim$0.11~km~s$^{-1}$ at 127~GHz). The achieved root mean square ($\sigma_{\mathrm{rms}}$) noise per 0.11~km~s$^{-1}$ channel is typically 0.028~K on the $T_{\rm mb}$ scale.  

Data reduction was carried out using the GILDAS/CLASS software \footnote{\url{https://www.iram.fr/IRAMFR/GILDAS/}}. Antenna temperatures ($T_A^\ast$) were converted to main-beam brightness temperature ($T_{\rm mb}$) using the beam efficiency $B_{\rm eff}$ interpolated from the IRAM 30m online tables, e.g., $B_{\rm eff} = 0.75$ at 127~GHz. The mapped area fully covers the dense core, with a beam size of $\sim$20\arcsec, allowing us to resolve the spatial distribution of CH$_2$NH across the carbon-chain shell of \ce{c-C3H2} peak and the central dust, HNCO and \ce{CH3OH} peak regions.

\begin{deluxetable}{lccccc}
\tablecaption{Detected \ce{CH2NH} line information.\label{tab:line_info}}
\tablehead{
  \colhead{Molecule} & \colhead{Transition} & \colhead{Frequency} &
  \colhead{$A_{\rm ij}$} & \colhead{$g_u$} & \colhead{$E_{\rm up}$}\\  
  \colhead{} &
  \colhead{} &
  \colhead{(MHz)} &
  \colhead{(10$^{-5}$ s$^{-1}$)} &
  \colhead{} &
  \colhead{(K)}
}
\startdata
\ce{CH2NH} & 2$_{0,2}$--1$_{0,1}$ & 127856.795 & 1.74 & 15 & 9.2 \\
\enddata
\tablecomments{Line data are adopted from the CDMS database:
\url{https://cdms.astro.uni-koeln.de/classic/}. Line frequencies are reported by \citet{Dore10}.}
\end{deluxetable}

\section{Analysis and Results}

The integrated intensity map of \ce{CH2NH} is shown in Fig. \ref{fig:mom0}, covering a velocity range of $V_{\mathrm{LSR}} (7.2)\pm0.3$~km~s$^{-1}$. The three molecular (family) peak positions and dust peak are shown. The emission of \ce{CH2NH} is rather extended across the core and generally follows the elongated core structure, with a modest enhancement close to the core center, with the brightest region only about 1.5–2 times above the surrounding emission level. The signal-to-noise map confirms this as a modest feature ($S/N\sim$6-8) located between the dust and \ce{HNCO} peaks. This pattern indicates that \ce{CH2NH} is widespread rather than concentrated at a specific position within the core. The average spectrum across the core (for $S/N$$>$3 pixels) is shown in Fig. \ref{fig:mom0} as well; the spectrum has a peak intensity of 0.17 K and seems to have a faint redshifted line wing around $\sim$8~km~s$^{-1}$. This extra small wing also appears in the average spectrum at the HNCO peak and results in an increment of linewidth (Fig. \ref{fig:4spectra}, see below).

The extracted spectrum at the four peaks are shown in Fig. \ref{fig:4spectra}, showing the averaged profile across a beam size. To place these spectra in context, we briefly recall the characteristics of the four molecular (family) peaks in L1544, discovered in \citet{Spezzano17}. The \ce{c-C3H2} peak marks the chemically young, carbon–chain–rich side of the core, where molecules such as \ce{C3H}, \ce{CCS} are also peaking. The dust peak corresponds to the region of highest molecular hydrogen column density where nitrogen-bearing species such as \ce{N2H+} typically show strong emission. The methanol peak highlights the location where several O-bearing species (e.g.\ \ce{CH3OH}, \ce{SO}, \ce{SO2}) are enhanced, likely originating from mild shocks of accretion flows (\citealt{Lin22}). Finally, the HNCO peak traces a distinct zone where \ce{HNCO}, propyne, and their deuterated isotopologues show their brightest emission (Giers et al., submitted). These chemically diverse environments provide useful reference points for diagnosing how the \ce{CH2NH} emission relates to the underlying physical and chemical structure of L1544. From Fig. \ref{fig:4spectra}, we can see that while there is no detection towards the methanol peak, a marginal detection towards the HNCO and dust peaks ($>2\sigma_{\mathrm{rms}}$) is observed with slightly broader linewidths than the spectrum that shows the strongest peak intensity at the \ce{c-C3H2} peak.


\begin{figure*}
\centering
\includegraphics[width=0.98\textwidth]{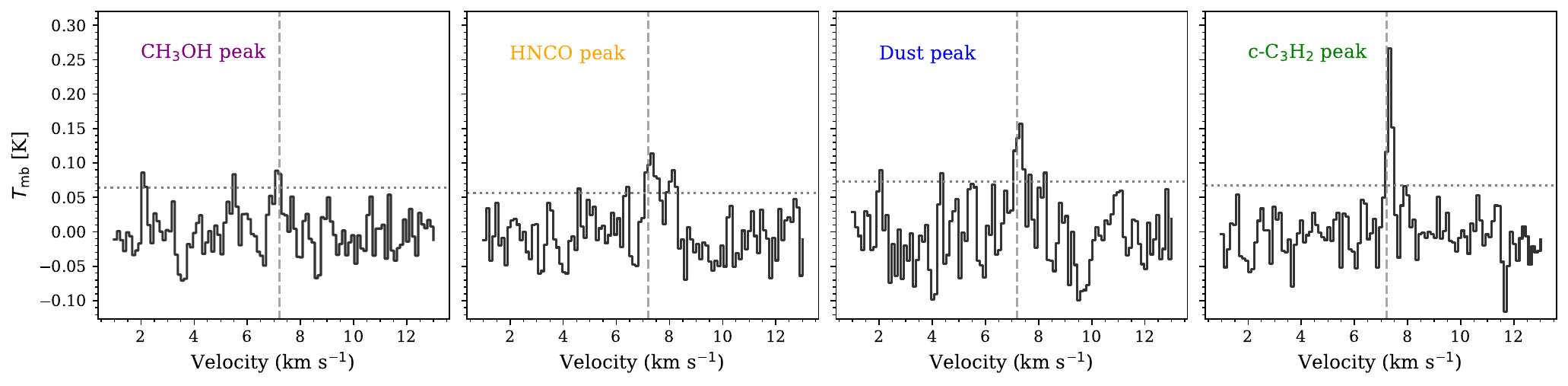}
\caption{The average \ce{CH2NH} 2$_{\rm{0,2}}$-1$_{\rm{0,1}}$ spectrum towards the three molecular peaks and the dust peak (Fig. \ref{fig:mom0}). The area used for spectrum averaging is one beam size, 20$''$. The vertical dashed line indicates $V_{\mathrm{lsr}}$ and the horizontal pointed line shows the 2$\sigma$ noise level. }
\label{fig:4spectra}
\end{figure*}

\begin{figure*}[htb]
\includegraphics[width=0.99\textwidth]{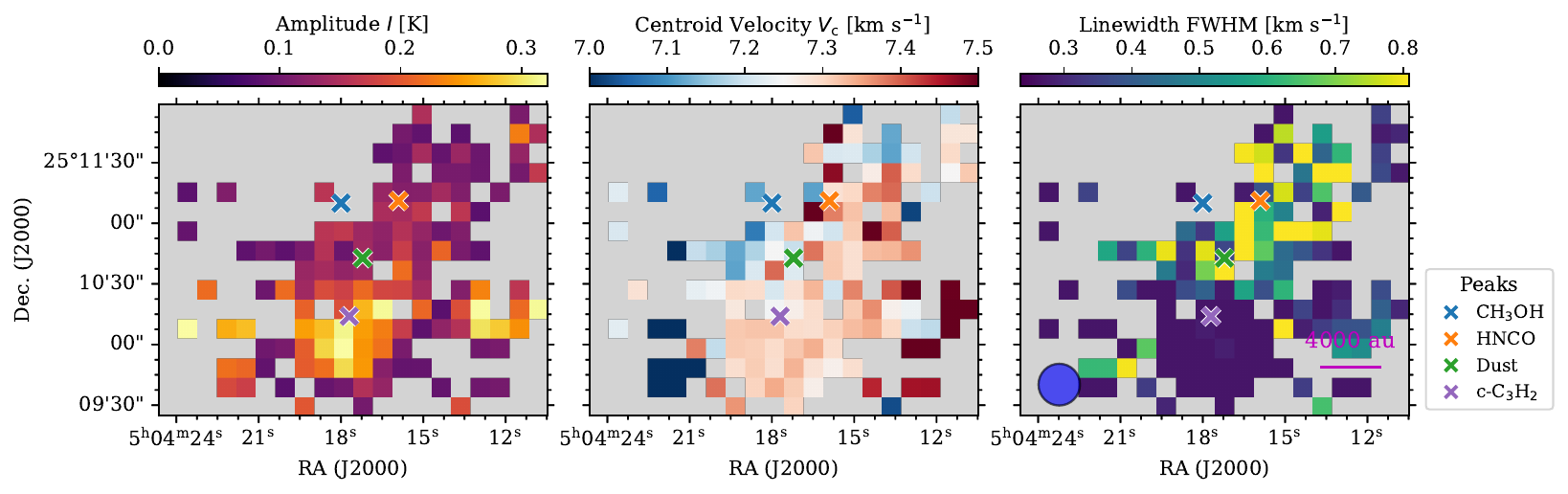}
\caption{The Gaussian fitted parameters of the \ce{CH2NH} line, showing the peak intensity, centroid velocity and velocity linewidth. The three molecular peaks and dust peak are marked. }
\label{fig:gauss}
\end{figure*}

We fit the line with a single-component Gaussian, deriving the peak intensity/amplitude $I$, centroid velocity $V_{\mathrm{c}}$, and velocity dispersion $\sigma_V$ (FWHM $= 2.355\sigma_V$). To estimate uncertainties, we perform a Monte Carlo analysis: for $n_{\rm MC}$ = 200 iterations, Gaussian noise based on the RMS of line-free channels is added to the spectrum, and each noisy realization is refitted with the same Gaussian model. The standard deviations of the resulting peak intensity, centroid, and width distributions provide the uncertainties $\sigma_{\rm{I}}$, $\sigma_{V_{\rm c}}$, and $\sigma_{\sigma_V}$. Afterward, we apply a quality check, masking pixels with low peak intensity ($I < 2.5 \sigma_{\rm rms}$) or large relative uncertainties ($\sigma_{\rm{I}} / I > 0.5$ or $\sigma_{\sigma_V} / \sigma_V > 0.5$), retaining only reliable measurements. The Gaussian fit results for the four spectra in Fig. \ref{fig:4spectra} are listed in Table \ref{tab:gauss_4spectra}.
The final parameter maps are shown in Fig.~\ref{fig:gauss}.

\begin{deluxetable}{lccc}
\tablecaption{The Gaussian fit parameters for average spectra across the core (for $S/N$$>$3 pixels), at dust peak and three molecular peaks (beam-averaged spectra).\label{tab:gauss_4spectra}}
 \tablehead{Location & Peak Intensity $I$ & $V_{c}$ & $\sigma_{V}$\\
         & (K) & (km~s$^{-1}$) &  (km~s$^{-1}$)}
         \startdata
         Core &0.15(0.01)&7.28(0.01)&0.15(0.02)\\
         Dust peak & 0.14(0.03) &7.21(0.06) &0.22(0.07) \\ 
         \ce{CH3OH} peak&0.08(0.01)&7.07(0.02)&0.18(0.02) \\
         \ce{HNCO} peak& 0.09(0.02)&7.41(0.11)&0.34(0.08) \\
         \ce{c-C3H2} peak & 0.25(0.01)&7.29(0.005)&0.11(1e-4)\\
         \enddata
\tablecomments{Numbers in brackets indicate fitting uncertainty (1$\sigma$) from Monte-Carlo simulations. $V_{c}$ stands for centroid velocity and $\sigma_{v}$ stands for velocity dispersion (FWHM/2.355).}
\end{deluxetable}

We find that the peak intensity of \ce{CH2NH} emission is located at the \ce{c-C3H2} emission peak, while the integrated intensity of \ce{CH2NH} peaks in-between the \ce{HNCO} and dust peak. This is because the velocity dispersion $\sigma_{\mathrm{V}}$ is larger at the northern region (Fig. \ref{fig:gauss}, see spectrum in Fig. \ref{fig:4spectra} as well) reaching $\sim$0.25-0.30~km~s$^{-1}$, while at the \ce{c-C3H2} peak the fitted $\sigma_V$ is similar to (or limited by) the channel width, 0.11~km~s$^{-1}$. The centroid velocity $V_{\mathrm{c}}$ appears to be shifted by $\sim$0.1-0.2~km~s$^{-1}$ compared to the system velocity, $V_{\mathrm{LSR}}$ = 7.2~km~s$^{-1}$ (e.g., \citealt{Tafalla98}).

\begin{figure*}[htb]
\includegraphics[width=0.9\textwidth]{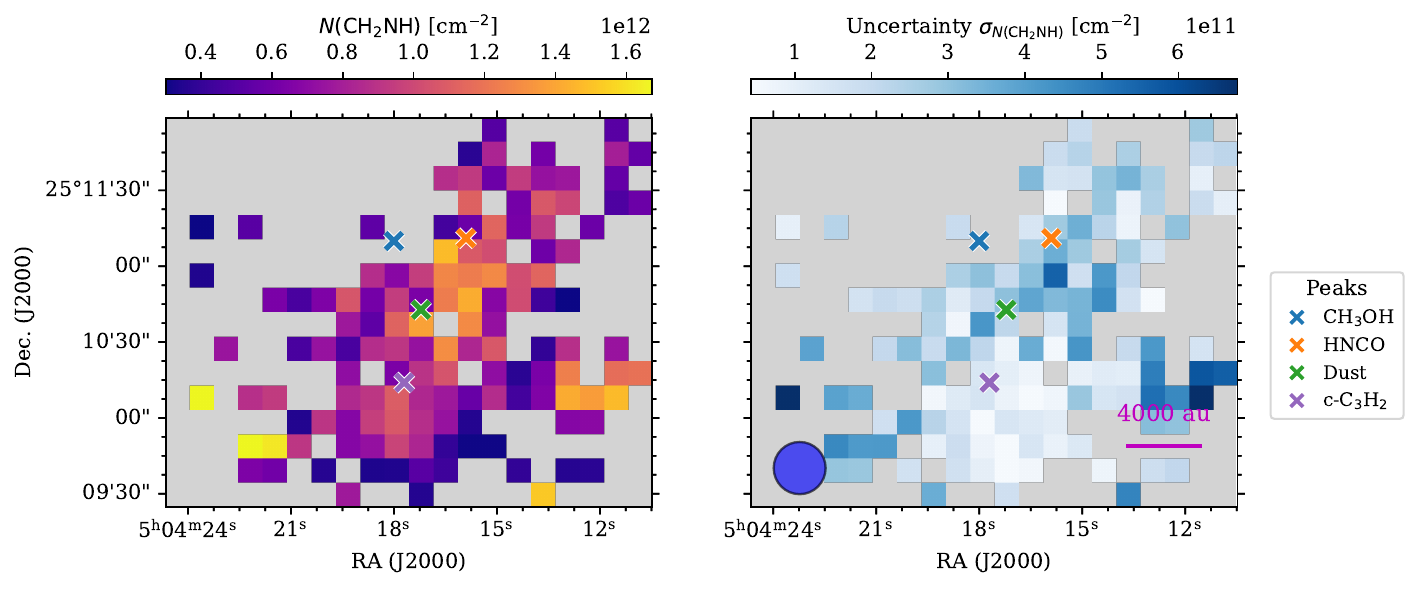}\\
\includegraphics[width=0.8\textwidth]{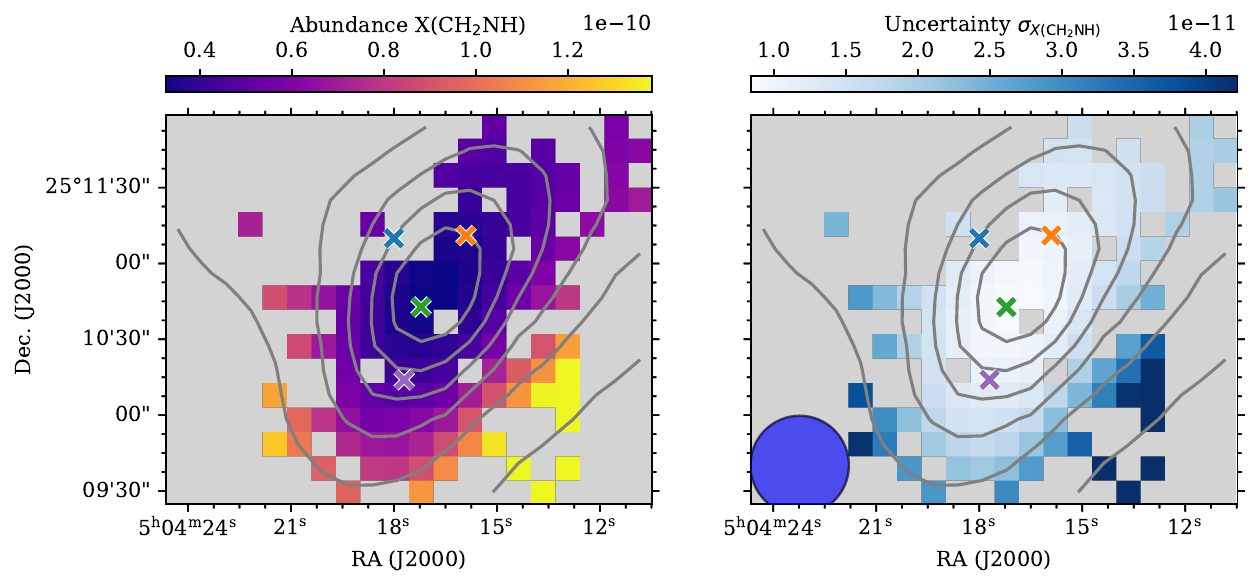}
\caption{Upper panels: The column density $N(\ce{CH2NH})$ and column density uncertainty map of L1544 assuming $T_{\mathrm{ex}}$ = 10 K. Lower panels: The \ce{CH2NH} abundance map and abundance uncertainty map, calculated with respect to $N_{\mathrm{H_2}}$ of 38$''$ angular resolution. Gray contours indicate molecular hydrogen column density ($N_{\mathrm{H2}}$) from 3.0$\times$10$^{20}$-2.7$\times$10$^{22}$~cm$^{-2}$ with six uniform intervals. The three molecular peaks and the dust peak are indicated with cross markers.}
\label{fig:nmol_x_map}
\end{figure*}

Since it appears that \ce{CH2NH} originates from an outer layer of L1544, we assume an excitation temperature of $T_{\mathrm{ex}}$ = 10 K and Local Thermodynamic Equilibrium (LTE) condition, and calculate the \ce{CH2NH} column density, $N(\ce{CH2NH})$ and the associated uncertainty map $\sigma_N(\ce{CH2NH})$ adopting the uncertainty in the integrated flux uncertainty (from the Gaussian fits). The column density of \ce{CH2NH} ranges from 0.4--1.4$\times$10$^{12}$~cm$^{-2}$ and peaks at the intermediate region of the dust peak and HNCO peak.
After convolving to the same beam as the $N_{\mathrm{H_2}}$ map and trimming further pixels with $N(\ce{CH2NH})$/$\sigma_N(\ce{CH2NH})$ $<$ 3, we derive the \ce{CH2NH} abundance $X(\ce{CH2NH})$ distribution and its uncertainty (Fig. \ref{fig:nmol_x_map}). The $X(\ce{CH2NH})$ across L1544 ranges from 0.3--1.2$\times$10$^{-10}$ and has a decreasing trend from the \ce{c-C3H2} peak to the dust peak and HNCO peak, from south to north.

\subsection{Comparison with other sources}


\ce{CH2NH} is detected across a wide range of astronomical environments, with its fractional abundance $X(\ce{CH2NH})$ varying by orders of magnitude. In general, in cold dark clouds and pre-stellar cores, \ce{CH2NH} seems extremely scarce (e.g., upper limits summarised in \citealt{Margules22}). In the starless core L183, the reported $X(\ce{CH2NH})$ is 8$\times$10$^{-10}$ \citep{Dickens97, Turner99}, while our measurements in L1544 yield lower $X(\ce{CH2NH})$ of 0.3–1.2$\times$10$^{-10}$ across the core. The low-mass protostar (hot corino) IRAS16293-2422B also shows a similar abundance of lower than $<$0.7$\times$10$^{-10}$ \citep{Ligerink18}, close to the starless and pre-stellar core values. These abundances are 10–100 times lower than typical hot-core values \citep{Suzuki16, Qin10, Crockett14}. The significantly higher \ce{CH2NH} abundance in massive cores is likely due to hydrogenation processes on dust grains starting from HCN; for example, values up to $\sim$7$\times$10$^{-8}$ have been reported \citep{Suzuki16}.

%
\begin{figure*}[htb]
    \includegraphics[width=0.98\textwidth]{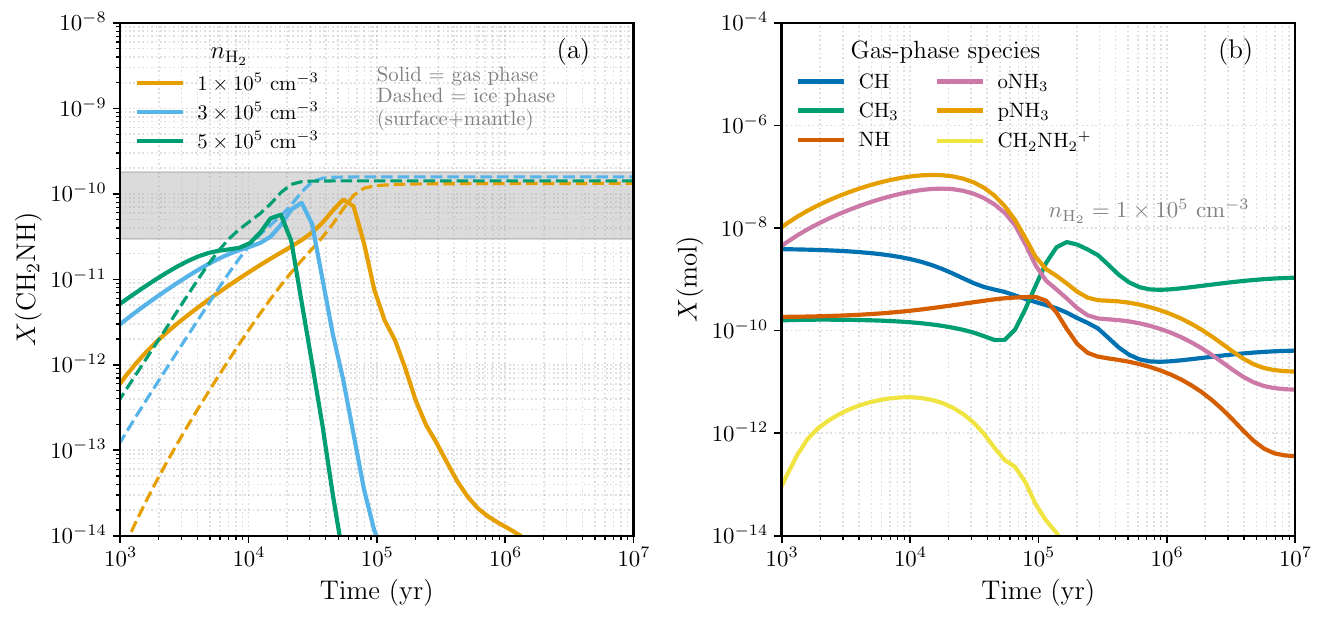}
  \caption{The abundance variation of \ce{CH2NH} (panel (a)) in the gas phase and on the grain surface. Three density setups (10$^{5}$~cm$^{-3}$, 3$\times$10$^{5}$~cm$^{-3}$, 5$\times$10$^{5}$~cm$^{-3}$) are shown. The shaded region indicates the observed $X(\ce{CH2NH})$ in L1544 above $N_{\mathrm{H_2}}$ of 5$\times$10$^{21}$~cm$^{-2}$. The variation of gas-phase abundance of related species (panel (b)) as a function of time, for density of 10$^{5}$~cm$^{-3}$.}
    \label{fig:chem_abu_gas_grain}
\end{figure*}

\subsection{Comparison with chemical models}

\begin{figure*}[htb]
\centering
     \includegraphics[width=0.92\textwidth]{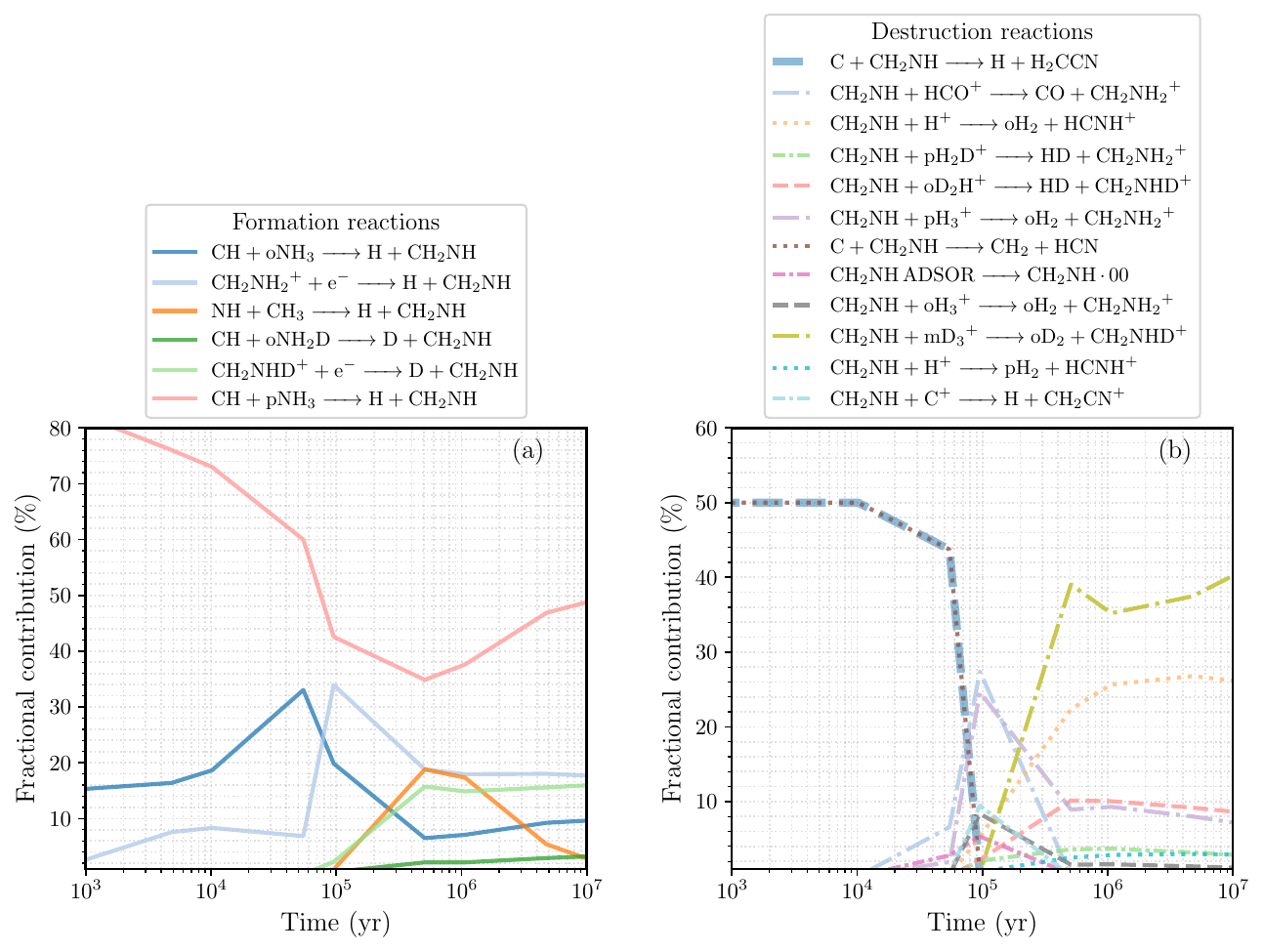}
     \caption{The dominant gas-phase \ce{CH2NH} formation (left panel) and destruction (right panel) pathways in the chemical model as a function of time, for a density of 10$^{5}$~cm$^{-3}$. 
     For each time epoch, we rank and select the top five dominant pathways to plot. The list of destruction reactions contains adsorption of \ce{CH2NH} onto the dust grains, as indicated by "\ce{CH2NH} ADSOR \ce{->} \ce{CH2NH}$\cdot$00", where \ce{CH2NH}$\cdot$00 denotes grain-surface \ce{CH2NH}.}
     \label{fig:reactions}
 \end{figure*}

To understand the chemistry that leads to the extended emission of \ce{CH2NH} that we observe in L1544, we compare the observed $X(\ce{CH2NH})$ with chemical model predictions. The simulated molecular abundances are derived with models at 10$^{5}$, 3$\times$10$^{5}$, and 5$\times$10$^{5}$~cm$^{-3}$ and 10 K using the state-of-the-art gas-grain chemical model pyRate described in \citet{Sipila15, Sipila19a}; We adopt the three-phase description of the chemical model, where the ice on grain surfaces is separated into an active surface layer and a chemically inert mantle beneath.
According to the physical model of L1544 (\citealt{KC10}), the average density at the three molecular peaks (0.02 pc from the dust peak) is $\sim$10$^{5}$~cm$^{-3}$ (see Fig.6 in \citealt{Lin22}), and the two higher densities represent gas layers closer to the dust peak.
We used the same initial abundances as in \citet{Sipila19a}, a constant cosmic-ray ionization rate $\zeta$ = 1.3$\times$10$^{17}$~s$^{-1}$, and a grain radius of 0.1~$\mu m$.
Figure~\ref{fig:chem_abu_gas_grain} shows the evolution of \ce{CH2NH} in the gas phase and on grain surfaces. Gas-phase \ce{CH2NH} peaks early and declines sharply after $10^{4}$–$10^{5}$ years, depending on density, marking it as an early-time chemical tracer, as in the case of \ce{c-C3H2}. Our observed $X(\ce{CH2NH})$ aligns with this peak. On grain surfaces, \ce{CH2NH} reaches a similar peak abundance during the pre-stellar phase and then plateaus, because the grain phase abundance is due to adsorption of gas-phase \ce{CH2NH} and no efficient grain-surface formation pathways are currently included in the model.

The observed $X(\ce{CH2NH})$ in L1544 peaks at the \ce{c-C3H2} molecular peak, a region showing abundant carbon-chain molecules influenced by the enhanced UV-driven radical chemistry (\citealt{Spezzano17}). It therefore seems that the UV-driven chemistry produces the radicals and ions needed to form \ce{CH2NH}. Indeed, examining the chemical models, we extract the dominant formation and destruction (adsorption) pathways of \ce{CH2NH} implemented in the chemical network. The primary gas-phase formation routes for \ce{CH2NH} are (Fig. \ref{fig:reactions}),

\begin{itemize}
    \item \ce{CH + NH3 -> H + CH2NH}: Methylidyne (\ce{CH}) reacts efficiently with ammonia. Laboratory and theoretical studies indicate this channel is barrierless and yields \ce{CH2NH} as the dominant product \citep{Suzuki16, Yang23}. Here we note that \ce{NH3} is not an early type chemical molecule, since its abundance in gas phase rises more rapidly in higher densities, so this route can be more important toward the core center. This is because \ce{NH3} needs \ce{N2} to form and the production of \ce{N2} is slow.

    \item \ce{CH2NH2+ (CH2NHD+) + e- -> CH2NH + H (D)}: Protonated methanimine recombines with electron. \ce{CH2NH2+} (mainly formed by \ce{CH3+ + NH3},  and also \ce{CH3+ + NH2} at early times) undergoes dissociative recombination to give \ce{CH2NH} (among other products). This is a major \ce{CH2NH} production channel \citep{Turner99, Yuen19}. 

    \item \ce{NH + CH3 -> H + CH2NH}: The reaction of the amidogen radical (\ce{NH}) with the methyl radical (\ce{CH3}) produces \ce{CH2NH} and a hydrogen atom. This neutral--neutral pathway has been invoked in models of hot cores (\citealt{Suzuki16}) and has been confirmed by chemical studies \citep{Joshi2022}. 
\end{itemize}

The main gas-phase destruction routes of \ce{CH2NH} are shown in Fig.~\ref{fig:reactions}(b). 
At early times before $t \sim 10^{5}$~yr, reaction with atomic carbon (to form either H + \ce{H2CCN} or \ce{CH2} + \ce{HCN}) completely dominate the destruction. In fact, the increased linewidth and the distribution of \ce{CH2NH} (integrated intensity peak) at the \ce{HNCO} peak resembles the case of \ce{CH3CCH} (Giers et al, submitted.), which is a molecule also mainly destroyed by C atoms. 
Around $t \sim 10^{5}$~yr, when the gas-phase abundance of \ce{CH2NH} peaks, its destruction is dominated by protonation reactions with abundant ions such as \ce{H+}, \ce{H3+}, and \ce{D2H+}, producing protonated methanimine (\ce{CH2NH2+}, \ce{CH2NHD+}) followed by dissociative recombination. 
Additional loss channels include reactions with \ce{C+} and \ce{HCO+}, forming \ce{CH2CN+} and \ce{CH2NH2+}, respectively. 

The spatial distribution of \ce{CH2NH} in L1544 reflects the interplay of these processes. Its abundance declines from south to north, peaking in the outer \ce{H2} layers rather than at the dust peak. There, moderate UV irradiation sustains radicals (\ce{CH}, \ce{NH}, \ce{CH3}) through photodissociation of small molecules (Fig. \ref{fig:chem_abu_gas_grain}), driving efficient gas-phase formation. Simultaneously, atomic carbon is partially depleted into more stable carbon-chain species (\ce{C2}, \ce{C3H}, \ce{c-C3H2}), limiting destruction via \ce{C + CH2NH -> H + H2CCN}. 
In the dense, well-shielded core near the dust peak, radical formation is suppressed and adsorption onto grains dominates, reducing gas-phase \ce{CH2NH}. This combination of active radical-driven formation, moderated destruction, and carbon-chain–rich chemistry traced by \ce{c-C3H2} explains the observed abundance gradient, underscoring the importance of moderate UV-driven photochemistry in the outer, less dense envelope \citep{Spezzano16, Spezzano17, Lin22}. 
The incorporation of \ce{CH2NH} into grain mantles in the dense core suggests that this molecule, a precursor to amino acids such as glycine, may be inherited during protostellar collapse. Quantifying this inheritance is critical for understanding the early chemical steps leading to complex organics in nascent planetary systems.

\section{Conclusion}
We present the first spatially resolved detection of the prebiotic molecule \ce{CH2NH} in the pre-stellar core L1544. The emission is extended across the core, with its peak intensity coinciding with the \ce{c-C3H2} region, while the integrated intensity shifts toward the zone between the HNCO and dust peaks, reflecting broader linewidths there. The measured abundances of $(0.3$–$1.2)\times10^{-10}$ reveal a clear chemical gradient from the carbon-chain–rich envelope to the dense, depleted center. Our comparison with chemical models indicates that \ce{CH2NH} forms efficiently through gas-phase neutral–neutral and ion–molecule reactions sustained by mild UV irradiation in the outer layers. This discovery demonstrates that key prebiotic N–C chemistry remains active even in the cold, quiescent phase preceding collapse, ensuring that organic precursors such as \ce{CH2NH} can be inherited by the next generation of forming stars and planets.


\begin{acknowledgements}
Y. Lin gratefully acknowledges the IRAM 30m staff for their kind help during the observation run. 
\end{acknowledgements}





%
\facilities{IRAM}


\bibliography{zref}{}
\bibliographystyle{aasjournalv7}



\end{document}